\documentstyle[emulateapj,apjfonts,psfig,times]{article}

\newcommand{\lapr}{\raisebox{-.6ex}{\mbox{
$\stackrel{<}{\mbox{\scriptsize$\sim$}}\:$}}}
\def\asca{{\sl ASCA}}
\def\ros{{\sl ROSAT}}
\def\chan{{\sl Chandra}}
\def\euv{{\sl EUVE}}
\def\sax{{\sl BeppoSAX}}
\def\nh20{n_{\rm H,20}}

\def\psr{PSR~J0437--4715}
\def\lx{L_{\rm X}}
\def\lb{L_{\rm bol}}
\def\rpc{R_{\rm pc}}
\def\tpc{T_{\rm pc}}
\def\rc{R_{\rm core}}
\def\tc{T_{\rm core}}
\def\rr{R_{\rm rim}}
\def\tr{T_{\rm rim}}
\def\Eb{E_{\rm b}}
\begin{document}
\lefthead{Zavlin et al.}
\righthead{\chan\ observations of \psr}
\title{
X-ray Radiation from the Millisecond Pulsar J0437--4715}
\author{
V.~E.~Zavlin\altaffilmark{1}\altaffiltext{1}{
Max-Planck-Institut f\"ur Extraterrestrische Physik, D-85740
Garching, Germany; zavlin@mpe.mpg.de},
G.~G.~Pavlov\altaffilmark{2}\altaffiltext{2}{
The Pennsylvania State University, 525 Davey Lab,
University Park, PA 16802, USA; pavlov@astro.psu.edu},
D.~Sanwal\altaffilmark{2}, 
R.~N.~Manchester\altaffilmark{3}\altaffiltext{3}{
Australia Telescope National Facility, CSIRO, P.~O. Box 76,
Epping, NSW 1710, Australia}, 
J.~Tr\"umper\altaffilmark{1},
J.~P.~Halpern\altaffilmark{4}\altaffiltext{4}{
Columbia University, 520 W.\ 120th Street, Mail Code 5230,
New York, NY 10027, USA}, 
and W.~Becker\altaffilmark{1}
}

\begin{abstract}

We report on spectral and timing observations of the nearest millisecond
pulsar J0437--4715 with the \chan\/ X-ray Observatory.  The pulsar
spectrum, detected up to 7~keV, cannot be described by a simple
one-component model.  We suggest that it consists of two components,
a nonthermal power-law spectrum generated in the pulsar magnetosphere,
with a photon index $\gamma\approx 2$, and a thermal spectrum emitted
by heated polar caps, with a temperature decreasing outwards from 2 MK
to 0.5 MK.  The lack of spectral features in the thermal component suggests
that the neutron star surface is covered by a hydrogen (or helium)
atmosphere.  The timing analysis shows one X-ray pulse per period, with
a pulsed fraction of about 40\% and the peak at the same pulse phase as
the radio peak.  No synchrotron pulsar-wind nebula is seen in X-rays.

\end{abstract}


\section{X-rays from millisecond pulsars: Thermal versus nonthermal}

Millisecond (recycled) radio pulsars are distinguished from ordinary pulsars
by their very short and stable periods,
$P\la 10$~ms, $\dot{P}\sim 10^{-21}-10^{-19}$~s~s$^{-1}$.
It is generally accepted that they are very old objects, with spin-down ages 
$\tau=P/2\dot{P}\sim 10^9-10^{10}$~yr and low surface magnetic fields 
$B\propto (P\dot{P})^{1/2}\sim 10^8-10^{10}$~G 
(e.g., Taylor, Manchester, \& Lyne 1993).
Similar to ordinary pulsars, a millisecond pulsar can emit 
nonthermal X-rays from its magnetosphere, with a hard power-law spectrum
and sharp pulsations. In addition to this nonthermal
radiation, thermal X-rays can be emitted
from the neutron star (NS) surface, provided the surface is hot
enough. According to the models of NS thermal evolution
(albeit rather uncertain at these old ages),
recycled pulsars are too cold (surface temperature
$T\la 0.1$~MK --- see, e.~g., Tsuruta 1998) to be detectable
in X-rays. However, their polar caps 
can be heated up to X-ray
temperatures by relativistic particles 
impinging onto the magnetic poles from the acceleration zones
in the magnetosphere.  The radio pulsar models
(e.g., Cheng \& Ruderman 1980; Arons 1981; Michel 1991;
Beskin, Gurevich, \& Istomin 1993) predict polar cap radii
$\rpc\sim (2\pi R^3/Pc)^{1/2}$ (where $R\approx 10$~km is
the NS radius), i.e., $\rpc\sim 1$--5~km for millisecond pulsars,
although different models predict quite different polar cap temperatures, in
the range of 1--10~MK. Detection of the polar cap thermal 
radiation would allow one
to discriminate between various models of radio pulsars, study the properties
of NS surface layers, and constrain the NS mass-to-radius ratio 
(Pavlov \& Zavlin 1997; Zavlin \& Pavlov 1998 [ZP98]). However, just as
in the case of ordinary pulsars, this radiation is detectable only if
it is not buried under stronger nonthermal radiation.  The current theoretical
models are not elaborate enough to predict in which (if any) of millisecond
pulsars the thermal component can be brighter than the nonthermal one (in
particular, both the thermal and nonthermal luminosities are expected to
increase with spin-down energy loss $\dot{E}$, perhaps with different rates).
Therefore, we have to rely upon the analysis of X-ray observations
to distinguish the thermal and nonthermal components.

The X-ray observatories \ros, \asca, 
and \sax\/ have detected 11
millisecond pulsars (nearly 1/3 of all X-ray-detected rotation-powered
pulsars --- see Becker \& Pavlov 2001 for a recent review).
Five of these pulsars are identified in X-rays only by positional
coincidence with the radio pulsars and, due to the low number
of recorded counts, provide only crude flux estimates.
The radiation from 3 pulsars --- B1821--24 (Saito et al.~1997), 
B1937+21 (Takahashi et al.~2001), and J0218+4232 (Mineo et al.~2000) ---
is clearly nonthermal: their power-law spectra, detected with \asca\/ and \sax\/
up to energies of 5--10~keV, are very hard, with photon indices $\gamma\sim 1$,
and their pulse profiles show sharp peaks.  Interestingly, these 3 pulsars are
characterized by particularly large $\dot{E}$ values, 
$\dot{E}=(2-20)\times 10^{35}$~erg~s$^{-1}$, 
and their magnetic fields at the light cylinder, 
$B_{\rm lc}=B(R/R_{\rm lc})^3\sim 10^6$~G,
are close to that of the Crab pulsar.  The case for the other 3 pulsars ---
J0437--4715 (Becker \& Tr\"umper 1993, 1999 [BT93, BT99]; ZP98),
J2124--3358 (BT99), and J0030+0451 (Becker et al.~2000) --- is less certain.
These pulsars show broad peaks of X-ray pulsations, but it does not
necessarily mean that their radiation is thermal because broad peaks
can be produced by nonthermal emission at some viewing angles.  High-quality
spectra have been recorded for 
the brightest of these pulsars, 
J0437--4715, but  their interpretation has been controversial --- e.g., ZP98
suggest that the radiation detected with \ros\/ and \euv\/ can be interpreted as
thermal radiation from hot polar caps, whereas BT99 argue that the radiation is
nonthermal (see \S2).  To resolve this controversy, the pulsar 
needed to be
observed at energies above the soft \ros\/ and \euv\/ bands ($E\ga 2$~keV), 
and with high spatial resolution to avoid contamination
of the pulsar emission by a nearby AGN which compromised the \asca\/
and \sax\/ data. 

The {\sl Chandra} X-ray Observatory provides both the superb spatial resolution
and high throughput at higher energies, together with timing capability.
In this paper we present the results of our observations of \psr\ with
\chan. We start from a summary of the previous results on \psr\ in \S2. 
The spectral and timing analyses of the \chan\/ data are presented in \S3
and \S4. Implication of the results are discussed in \S5.

\section{Previous X-ray observations of \psr}

At a distance $d=139\pm 3$~pc (van Straten et al.~2001),
\psr\ is the nearest and
brightest X-ray millisecond pulsar known
at both radio and X-ray wavelengths.
 It was discovered by Johnston
et al.~(1993) during the Parkes Southern radio pulsar survey.
It has a spin period $P=5.76$ ms,
and, after correcting the observed period derivative for the kinematic term
due to its large proper motion,
a characteristic age $\tau\simeq 4.9$ Gyr,
magnetic field $B\sim 3\times 10^8$ G,
and rotation energy loss rate $\dot{E}\sim 
3.8\times 10^{33}$ erg~s$^{-1}$.
It is in a 5.74~d binary
orbit with a white dwarf companion of a low mass $M\simeq 0.2 M_\odot$. 
Optical observations in H$_\alpha$ have revealed a bow-shock pulsar-wind
nebula (PWN), caused by the supersonic motion of the pulsar through
the interstellar medium,
with the bow-shock apex at about $10''$ south-east of the pulsar,
in the direction of the pulsar's proper motion
(Bell, Bailes, \& Bessell 1993; Bell et al.~1995). 

\psr\ was the first millisecond pulsar detected in X-rays --- BT93 observed it
with the \ros\/ Position Sensitive Proportional Counter (PSPC) in 1992 and
discovered X-ray pulsations with a single broad pulse and a pulsed fraction
$f_{\rm p}\sim 30\%$ in the 0.1--2.4~keV energy range. They also found, and
Halpern, Martin, \& Marshall (1996; HMM96 hereafter) later confirmed, that
$f_{\rm p}$ apparently increases with photon energy $E$ and reaches about 50\%
at $E= 0.6-1.1$~keV.  However, BT99 found no energy dependence of $f_{\rm p}$
from the PSPC observation of 1994. On the other hand, ZP98 analyzed the same
data and claimed that the energy dependence of the pulse profile is 
statistically
significant.

The pulsar spectrum is incompatible
with a simple blackbody (BB) model even
in the narrow PSPC band (BT93), but it can be fitted with a power-law
(PL) model, with a photon index $\gamma=2.2$--2.5 (BT93; HMM96; ZP98; BT99),
indicating that the soft X-ray radiation might be nonthermal.  However, the
probable energy dependence of the pulsed fraction makes this interpretation
questionable.  In addition, the combined analysis of the PSPC and \euv\/
Deep Survey Instrument (DSI) data show that the PL fit is only marginally
acceptable (HMM96; ZP98). 

BT93 suggested that the pulsar spectrum consists of two components
--- in addition to a PL ($\gamma\approx 2.9$), representing magnetospheric
emission, there is a BB component with $T_{\rm bb}\sim 1.7$~MK emitted
from a small area of 0.05~km$^2$, 
implying the existence of polar cap(s) on the NS
surface.  An alternative two-component model was explored by ZP98.  They
suggest that the temperature distribution in a polar cap should be nonuniform 
because the heat, released by decelerating magnetospheric particles
in subphotospheric layers, propagates along the surface out from the small
polar area where the energy is deposited, which results in a larger heated
area with the temperature decreasing outwards.  Applying the hydrogen NS
atmosphere models of Zavlin, Pavlov, \& Shibanov (1996), ZP98 show that the
entire soft X-ray spectrum and the pulse profiles detected with \ros\/ and
\euv\/ can be interpreted as thermal emission 
from two opposite polar caps where the
temperature drops from $\tc=1-2$~MK in a central cap (``core'') of radius
$\rc=
0.2-0.5$~km to $\tr=0.3-0.5$~MK in the annulus (``rim'') with an outer
radius $\rr= 2-5$~km.  The bolometric luminosity of the polar caps\footnote{
The values of polar cap radii and luminosities given in ZP98 
should be multiplied
by $\pi^{1/2}$ and $\pi$, respectively, because they used an erroneous 
normalization factor.  This error does not affect their qualitative
conclusions.}, $\lb= (2-3)\times 10^{30}$~erg~s$^{-1}$, is consistent
with the prediction of the slot-gap model of radio pulsars developed by
Arons (1981).  The hydrogen column density towards the pulsar,
$\nh20=n_{\rm H}/(10^{20}~{\rm cm}^{-2})=0.1-0.3$, 
obtained by ZP98,
is consistent with the properties of the interstellar medium inferred from
observations of other objects in the vicinity of the pulsar.

The \ros\/ energy band, 0.1--2.4~keV, is too soft to discriminate between
different two-component models --- e.g., BT99 have shown that equally good
fits of the \ros\/ PSPC spectrum can be obtained with a two-component BB+BB
model or a broken PL.  The pulsar has been observed at higher energies with
\asca.  Kawai, Tamura, \& Saito (1998) report that the \asca\/ spectrum is 
more consistent with a BB model with $T_{\rm bb}\approx 3$~MK than with a
PL model.  However, because of poor spatial resolution of \asca, the pulsar
spectrum was heavily contaminated by the radiation from a neighboring Seyfert
galaxy, that hampered the spectral analysis. Therefore, to investigate the
pulsar spectrum at higher energies and measure its pulse profile with high
accuracy, we proposed to observe \psr\ with \chan.  The results of the
spectral, timing, and spatial analyses are presented below.

\section{ACIS Observation and spectral analysis}

\psr\ was observed with the Advanced CCD Imaging Spectrometer
(ACIS; Garmire et al.~1992; Bautz et al.~1998) on 2000 May 29--30,
with a 25.7~ks effective exposure. It was imaged on the back-illuminated
chip S3 of the ACIS spectroscopic array.  To mitigate the pile-up
effect\footnote{See {\sl Chandra} Proposers' Observatory Guide (POG);
{\tt http://asc.harvard.edu/udocs/docs/docs.html}}, the pulsar 
was offset by $3\farcm9$ from the optical axis and observed using a 1/8
subarray, with a 0.4~s frame time.  We extracted source-plus-background
counts from a $5''$-radius circle centered on the pulsar image, which
includes about 98\% of pulsar counts.  To measure the background, we
extracted counts from several regions at distances of at least $10''$
from the pulsar, where the pulsar contribution is negligible.  The
background, 0.020 counts~ks$^{-1}$ arcsec$^{-2}$, is rather uniform
throughout the observed $8'\times 1'$ sky area --- in particular, we
see no structures which could be interpreted as a PWN associated with
the bow shock observed in H$_\alpha$.  The source count rate is
$306\pm 4$~counts~ks$^{-1}$ (0.12 counts per frame), for the standard 
grades 02346, in the 0.2--7.0~keV range (where the source prevails over
the background).

To examine the effect of pile-up on the pulsar spectrum, we extracted the
count-rate spectra from a $1''$-radius circle centered at the pulsar (which
contains 56\% of source counts) and from an annulus with $1''$ and $5''$
radii.  We compared the spectral shapes using a $\chi^2$ test and found them
to be the same with a probability of 92\%--95\% (depending on binning chosen).
Since the annulus spectrum should not be affected by pile-up, we conclude
that the effect of pile-up on the whole pulsar spectrum is negligible.

To search for possible spectral features in the ACIS spectrum, we used a
method applied by Pavlov et al.~(2001) to the dispersed 
X-ray spectrum of the Vela pulsar, obtained with the {\sl Chandra} Low Energy
Transmission Grating.  
We first binned the original detector channels of a 5~eV
width in bins of an 80 eV width (comparable with the intrinsic detector
resolution of $\sim 100$~eV).  Then, we combined 4--5 sequential binned
channels in groups of 320--400~eV widths to estimate the deviation in the
number of source counts in each binned channel of a given group from the
mean value in the group.  The maximum deviation does not exceed a
1.8~$\sigma$ level, which allows us to conclude that the ACIS spectrum of
\psr\ reveals no significant spectral features.  For further analysis, we
binned the pulsar spectrum in 75 energy bins in the 0.2--7.0~keV range.
For fitting the ACIS data, we used CCD responses
generated with the CIAO\,2.1 software and CALDB\,2.7 calibration
files.

To investigate as broad energy range as possible, we analyzed the ACIS
spectrum together with the \ros\/ PSPC spectra 
observed on 1992 September
20--21 (5.9~ks exposure) and 1994 July 2--6 (9.9~ks exposure).
(A detailed description of the PSPC data and their analysis can be found in
ZP98.)
Our attempts to simultaneously fit the ACIS and PSPC spectra immediately
showed that they disagree with each other at energies below 0.6~keV --- 
models which fit the ACIS and PSPC spectra at $E>0.6$~keV are well above
the observed ACIS data points at lower energies
(particularly in a 0.3--0.5 keV range).  We have encountered
the same problem with the Vela pulsar (Pavlov et al.~2001) and
PSR~1055--52 (Sanwal et al.~2000) --- the ACIS and PSPC spectra of these pulsars
are in good agreement only if we discard ACIS events below 
$\simeq 0.6$ keV.
This inconsistency is apparently
associated with errors in the ACIS response at low
energies, so we have to
use the ACIS data only in the 0.6--7.0~keV range (63\% of the source
counts in 50 energy bins).

\subsection{Single power law fit}

Fitting the separate ACIS spectrum in the 0.6--7 keV range with a PL model
yields a best-fit photon index $\gamma=4.1$
(min $\chi^2_\nu=61.17/47=1.30$),
drastically different from
that obtained for the PSPC spectrum ($\gamma= 2.4$).  Not surprisingly,
a single PL does not fit these spectra together
(${\rm min}~\chi^2_\nu=366.8/96=3.82$), clearly indicating that more
components are needed to get a satisfactory combined fit.

\subsection{Power law plus single-temperature polar caps}

It has been shown by ZP98 that even the narrow-band PSPC spectrum
cannot be interpreted in terms of a single-temperature polar cap radiation
--- there is always an excess of observed counts over the model
spectrum at higher energies, even for the hydrogen-atmosphere model
that provides the hardest spectrum.

\subsubsection{Hydrogen polar caps}
To explore the possibility that the observed X-ray emission consists of
thermal (polar cap) and nonthermal (magnetospheric) components, we start with a
two-component model: PL plus two polar caps covered with 
hydrogen or helium\footnote{
The difference between the hydrogen and helium spectral models is very small
in the X-ray range, at the temperatures of interest.}.  
We assume
the gravitational redshift parameter $g_r\equiv (1-2GM/Rc^2)^{1/2} =0.769$,
which corresponds to the NS radius $R=10$ km at $M=1.4 M_\odot$. 
To model the polar cap component, we should specify
some values for $\zeta$, the angle between the
line of sight and rotation axis, and $\alpha$, the angle between the
magnetic and rotation axes.
Measuring these angles 
from the swing of the radio
polarization position angle at the pulse center is 
difficult for this pulsar because of 
the complex position angle variations, apparently
including polarization mode transitions
(Navarro et al.\ 1997). However, it is reasonable to
assume that the rotation axis of the pulsar is close the orbital
momentum axis whose inclination angle, $\approx 43^\circ$, was
determined by van Straten et al.~(2001). On the other hand,
the radio pulse morphology
and polarization strongly suggest that the magnetic inclination
$\alpha$ is very close to $\zeta$, probably the difference does not
exceed $2^\circ$. We choose $\zeta=\alpha=45^\circ$ for our models
of polar cap radiation. We have checked that variation of the angles by
$\pm 10^\circ$ around these values makes no appreciable effect 
on the spectral parameters, and changes of the inferred radii of the polar caps
do not exceed 15\%.

This model provides an acceptable fit to the 
PSPC and ACIS data (see Fig.~1, left panels),
with a minimum $\chi^2_\nu=1.25$ (94 d.o.f)
and best-fit parameters $\nh20=0.6\pm 0.2$,
$\tpc=1.80\pm 0.05$~MK, 
$\rpc=0.15\pm 0.02$ km (at $d=140$ pc), and
$\gamma=2.8\pm 0.1$ (the uncertainties here and below are given at a 68\% 
level; the polar cap parameters are given as measured at the NS surface; the
gravitational effects are taken into account to calculate the model spectra
as seen by a distant observer 
--- see Zavlin, Shibanov, \& Pavlov 1995).  
The bolometric luminosity of two polar caps is
$\lb=(0.9\pm 0.2)\times 10^{30}$~erg~s$^{-1} \sim 0.2\times 10^{-3}\dot{E}$,
and the luminosity of the nonthermal component in the 0.1--10~keV range is
$\lx=(3.1\pm 0.4)\times 10^{30}$~erg~s$^{-1} \sim 0.8\times 10^{-3}\dot{E}$.
The thermal component prevails in the 0.6--2.7~keV band; the PL component,
dominating at higher and lower energies, determines the estimated value of
$n_{\rm H}$ and gives the main contribution to the \euv\/ DSI flux.  Comparison
with the \euv\/ DSI data shows that this 
model is only marginally 
acceptable,
similar to the case of the single PL fit to the PSPC spectrum, described by
HMM96 and ZP98. Also, the $n_{\rm H}$ values allowed by the fit to the
DSI, PSPC and ACIS data still exceed the values $\nh20=0.1-0.3$ derived 
from independent measurements of the hydrogen column density towards
objects in the vicinity of \psr\ (see ZP98). 

\subsubsection{Blackbody polar caps}
If we assume a standard blackbody model for the polar cap radiation,
we obtain qualitatively
similar results, with a higher temperature of the thermal component,
$T_{\rm bb}\simeq 2.9$~MK,
and a much smaller radius of emitting area,
$R_{\rm bb}\simeq 0.04$~km. 
Spectral fits with the simplified blackbody model consistently produce
higher temperatures and smaller sizes because X-ray spectra emerging
from light-element atmospheres are harder than blackbody spectra
(Zavlin et al.~1996).

\subsubsection{Iron polar caps}
Although the pulsar spectral fits are very comfortable with featureless
models, it is worthwhile to check a hypothesis that the surface
layers of polar caps are comprised of heavier chemical elements, e.~g., iron.
Using NS iron atmosphere models (Rajagopal \& Romani 1996; 
Zavlin et al.~1996\footnote{The iron atmosphere
spectra presented by Zavlin et al.~(1996) were computed with an error in
interpolating the opacity tables. We are grateful to R.~Romani who brought
our attention to this issue. The error has been corrected in the models used
in this paper. An example of a corrected model spectrum has been presented by
Pavlov \& Zavlin (2000).}) for the thermal (polar cap) component, 
we found that
the single PSPC spectrum fits well with a power law plus iron atmosphere model,
whereas the ACIS data of much better spectral resolution
are inconsistent with this model, particularly around the (redshifted)
L and K absorption features in the iron atmosphere spectrum
(at energies of 0.85 keV and 4.9 keV for $g_r=0.769$ ---
see right panels in Fig.~1).
Varying the gravitational redshift shifts the feature 
energies in the model spectrum,
but the features remain too strong to be consistent with 
the smooth observed spectrum
at any allowed values of $g_r$.

\subsection{Power law plus polar caps with  nonuniform temperature}

ZP98 showed that a 
polar cap model with a nonuniform temperature distribution along
the 
cap's surface fits the PSPC spectrum without invoking a nonthermal component.
They suggested a two-step approximation for the polar cap 
temperature distribution,
a hot ``core'' plus a colder ``rim'' (see \S2 and left panels of Fig.~2).
Applying this model to the ACIS data shows that the observed spectrum
significantly exceeds the model at $E>2$~keV (the upper limit of the PSPC
energy range) and obviously requires one more component.  A natural choice
for the additional component is a PL.  Adding the nonthermal component greatly
improves the fit (right panels in Fig.~2) reducing $\chi^2_\nu$ from 2.39 to
1.13.  This model has seven fitting parameters, some of them are strongly
correlated (ZP98).  The correlation is particularly strong between $n_{\rm H}$
and $\rr$. Since the best-fit values of the hydrogen column density are well
consistent with the values $\nh20=0.1-0.3$ obtained from the independent
estimates, we chose to fix the column density 
to weaken the correlations between the other six fitting parameters.
For $\nh20=0.2$, we obtain
$\tc=2.1^{+0.2}_{-0.3}$~MK, $\tr=0.54^{+0.06}_{-0.10}$~MK, 
$\rc=0.12^{+0.04}_{0.02}$ km, $\rr=2.0^{+0.3}_{-0.2}$ km,
and $\gamma=2.2^{+0.3}_{-0.6}$.
Varying the hydrogen column density within the
$\nh20=0.1-0.3$ range results in a minor change ($<7\%$)
in the values of $\rr$ and $\rc$ (at the other parameters fixed).
In this model the bolometric luminosity of two 
polar caps
is
$\lb=(2.3\pm 0.6)\times 10^{30}$~erg~s$^{-1}\sim 0.6\times 10^{-3}\dot{E}$,
and the PL luminosity in the 0.1--10~keV range is
$\lx=(0.7\pm 0.3)\times 10^{30}$~erg~s$^{-1}$
$\sim 0.2\times 10^{-3}\dot{E}$.
Note that this model is in good agreement with the \euv\/ DSI data.

One may assume that the additional component to fit the hard tail of the
ACIS spectrum is also a thermal one, instead of the PL.  However, this thermal
component has to be of a very high temperature, $T\sim 12-15$~MK, emitted from
a very small area of 1--2~m radius.  Although we are not aware of reliable
calculations of temperature distributions around pulsar's magnetic poles, we
believe 
that such a steep gradient in the temperature
distribution is unlikely.

\subsection{Broken power law}
Although the 
power-law plus polar-cap
models fit the PSPC+ACIS spectra satisfactorily, one
cannot rule out {\it a priori} that the (purely nonthermal) pulsar
spectrum bends down somewhere in the X-ray band, which might be caused, e.g.,
by a deficit of high-energy electrons in the pulsar's magnetosphere.  To test
this hypothesis, we applied a broken PL model to the PSPC and ACIS spectra and
obtained an acceptable fit ($\chi^2_\nu=1.31$) with $\nh20=0.3^{+0.5}_{-0.2}$
and photon indices $\gamma_1=2.0\pm 0.2$ and $\gamma_2=3.6\pm 0.1$, below and
above the break energy $\Eb=1.1\pm 0.1$~keV, respectively. However, the \euv\/ 
DSI flux calculated with the best-fit parameters exceeds the observed DSI flux
by about $3\sigma$ (if one adopts a 15\% systematic uncertainty in the DSI 
flux).
Including the DSI data in the combined analysis shows that the broken PL model  
can be accepted at the same low-confidence level as in the case of the PL plus
single-temperature polar cap model fit,  and it requires $\nh20>0.5$.

Broken PL spectra have been observed from young ordinary pulsars, but with
a much higher break energy, in the GeV range.  However, no breaks have been
observed in the spectra of other millisecond pulsars, and the (single) PL
slopes measured in the three millisecond pulsars with certainly nonthermal
radiation are more gradual than the slopes of the broken PL components in
\psr.  Therefore, although we cannot completely reject the broken PL 
interpretation, we consider it less plausible than the 
power-law plus thermal polar-cap model.

\section{HRC Observations and Timing Analysis}
\psr\ was observed with the imaging array of the High Resolution Camera 
(HRC-I; Murray et al.~1997) on 2000 February 13, with a 18.9~ks 
effective exposure.  The pulsar was detected with a source count rate of
$174\pm 4$ counts~ks$^{-1}$ (estimated in a $3''$-radius circle centered
at the pulsar's position). 
The image of the pulsar and its surroundings is shown in Figure 3.
The surface brightness of the background in the
pulsar vicinity, 0.015~counts~ks$^{-1}$~arcsec$^{-2}$ is so low that its
contribution is negligible in the source aperture, and the count rate error
is determined solely by the source photon statistics.  Figure~4 shows the
extracted radial distribution of the (energy-integrated) encircled source
count fraction. The distribution is well consistent with that expected for
a point source.

Our attempts to detect pulsations in these data failed.  The reason of this
failure was found later --- a wiring error in the HRC causes the time of
an event to be associated not with that event, but with the following event
at the HRC front end, which most likely is a background particle event
rejected by the anti-coincidence shield and not telemetered to the ground
(see POG, \S7.10; Tennant et al.\ 2001).  
As a result, most of the true event times are lost, and  
the HRC timing accuracy degrades from the planned 16~$\mu$s to 
the mean time between events at the HRC front end, typically about 4~ms.

To circumvent this problem, the HRC team worked out a special operating mode
which employs the HRC-S instrument.  This mode uses only the central segment
of the HRC-S, with two outer segments disabled, which lowers the overall
counting rate below the telemetry saturation limit, even with all on-board
vetoes disabled.  As a result, all events are telemetered and can be assigned
correct times by shifting the time-tags in ground processing.

The pulsar was re-observed in this new mode on 2000 October~6
(19.6~ks effective exposure).  We measured the source count rate
of $152\pm 6$ counts~ks$^{-1}$ using the same aperture as for
the HRC-I data.  To perform the timing analysis, we first extracted
the ``arrival times'' (as given in Level 2 data file) of  2,795 
events from a $1''$-radius circle centered at the pulsar (93\% of the total
number of source events).  Then, for each of the selected events, we found the
time of the following event in the Level 1 data file (which, in addition to
photon events, contains time-tagged particle events) and assigned this later
time to the selected source event.  These corrected times were converted to 
equivalent arrival times at the solar system barycenter (SSB) in 
the barycentric dynamical time (TDB) system with the {\tt axBary}
tool\footnote{
See: {\tt http://asc.harvard.edu/ciao2.2/test/ahelp/axbary.html}}
using the solar system ephemeris DE200 (Standish 1982).

To correct the TDB arrival times for the effect of the orbital Doppler shift
(Taylor \& Weisberg 1989) and extract the light curve, we used the pulsar and
binary orbit ephemerides given in Table 1.
The values of $Z_n^2$ statistics (Buccheri et al.~1983) calculated
with the X-ray arrival times
at the radio ephemeris parameters
are $Z_1^2=217$, $Z_2^2=258$,
$Z_3^2=259$.  The difference between the radio period and that 
corresponding to the $Z_2^2$ maximum, $\delta P=3\times 10^{-12}$~s
(or $\delta F=\delta P/P^2=1\times 10^{-7}$~Hz), is negligible
for the timing analysis of the X-ray data collected in a 20~ks time span. 
Note that $\delta{P}$ is much smaller 
than the 1\,$\sigma$ uncertainty (given by the Bayesian method ---
see Zavlin et al.~[2000] for details), $1\times 10^{-10}$ s,
of the period found in the HRC-S data. The small value of $\delta P$ and
the large $Z_n^2$ values prove unambiguously that the new HRC-S timing mode
is fully capable of detecting millisecond pulsations.  (The largest $Z_n^2$
values achieved in the \ros\/ observations of this pulsar are 117 and 125 for
$n=1$ and $n=2$, respectively).

The upper panel of Figure~5 presents the extracted HRC-S (energy-integrated) 
pulse profile of \psr. Using the radio ephemeris given in Table 1
and radio observations at frequencies close to 1400 MHz, made using
the Parkes Telescope and the Caltech correlator (cf. Sandhu et al.~1997),
which were contemporaneous with the X-ray observations, 
we were able to find the relation between the radio and X-ray
phases --- $\phi=0$ in this figure corresponds to the maximum of
the radio peak observed at 1420 MHz, dedispersed
to infinite frequency using the DM in Table~1. The epoch
of $\phi=0$\  is
     $51\,823.7311148130996$ MJD [UTC at geocenter].
Using {\sl TEMPO}\footnote{TEMPO is a program for the analysis of pulsar
timing data maintained and distributed by Princeton University and the
Australia Telescope National Facility. It is available at
{\tt http://www.atnf.csiro.au/research/pulsar/timing/tempo/}},
we determined the phase offsets of the X-ray peaks for five
sub-intervals within the 20-ks exposure. The corresponding time
offsets vary from  --6 $\mu$s to +37 $\mu$s,
with a mean value of +16 $\mu$s (0.3$\%$ of the pulsar period), 
equal to the intrinsic HRC time resolution.
Therefore, we conclude that the phases of the X-ray and radio pulses 
are virtually the same --- their difference does not exceed the errors
of the \chan\/ absolute time\footnote{
The accuracy of the \chan\/ absolute time was estimated by 
Tennant et al.\ (2001) from the comparison
of the \chan\/ HRC-I and {\sl RXTE} timing of the Crab pulsar:
$t_{\rm Chandra} - t_{\rm RXTE} = -200\pm 100$ $\mu$s.}.
It should be noted that this is the first determination of the
absolute phase of X-ray pulsations of \psr\ --- errors in absolute
timing of the \ros\/ data were at a 1--2 ms level (see HMM96 and
BT99 for discussion), preventing the absolute phase determination.

For comparison, we also show in Figure 5 the pulse profiles
obtained from the \ros\/ PSPC and HRI observations of 1994 (with phases
chosen to approximately co-align the \ros\/ peaks with 
\chan\/ peak). The pulsed fraction \footnote{
Conventionally defined as the 
ratio of the number of counts 
above the minimum of the light curve to the total number of counts.}
of the HRC-S light curve
is somewhat larger than those in the \ros\/ data.  Since the HRC-S is
more sensitive at higher energies than the \ros\/ detectors, this may
indicate that $f_{\rm p}$ increases with $E$, as predicted by the
hydrogen polar cap model.  In addition, the pulse in the HRC-S light curve
looks slightly narrower than the PSPC and HRI pulses.  This also
could be explained by the properties of the thermal radiation
emitted from a hydrogen atmosphere --- the radiation intensities
are more beamed at higher $E$ (Zavlin et al.~1996).

\section{Discussion}

The \chan\/ observations of \psr\ have allowed us to perform the spatial,
timing and spectral analyses of the new data collected with high angular
and spectral resolution in an extended energy range.

The principal new result of the ACIS observation is the measurement of the
pulsar's X-ray spectrum at higher energies, up to 7~keV.  Among the formally
acceptable fits of the combined \ros\/ PSPC and \chan\/ ACIS spectra, only the
broken PL model corresponds to a purely nonthermal pulsar radiation in the
X-ray band.  However, the break energy of about 1.1 keV is 5--6 orders of
magnitude lower than those observed in other pulsars.  Such a difference
looks too large to be explained by a lower magnetic field in the radiating
region --- it would also require a lower break energy in the spectrum of
radiating electrons/positrons, compared to ordinary pulsars.  Moreover, 
extrapolation of the broken PL model to the optical $B$ and $V$ bands
(assuming the same slope as in the X-ray range below 1.1~keV)
predicts optical magnitudes, 
$m_B=19.6\pm 1.4$ and
$m_V=19.1\pm 1.4$ (for extinction coefficients
$A_B=0.3$, $A_V=0.2$ --- Danziger, Baade, \& della Valle 1993),
much brighter than those detected from the white dwarf companion,
$m_B=22.1\pm 0.1$ and 
$m_V=20.9\pm 0.1$ (Bailyn 1993; Danziger et al.~1993; Bell et al.~1993).  Thus,
we do not consider the broken PL model to be a plausible interpretation.

The alternative model involves {\sl two components} of different origin ---
thermal and nonthermal.  The nonthermal component originates in the pulsar
magnetosphere\footnote{Another source of unresolvable
nonthermal radiation
could be the shocked pulsar wind near the white dwarf companion
(e.g., Arons \& Tavani 1993). However, 
at the distance
of $a_p=1.5\times 10^{11}$ cm from the pulsar
(van Straten et al.\ 2001), the companion intercepts
only a fraction $\sim 6\times 10^{-4}$ of the wind
(assuming the wind is approximately isotropic),
too small to explain the observed nonthermal luminosity.},
whereas the thermal component is emitted from hot polar caps on the
NS surface.  
Depending on assumption about the polar cap temperature distribution,
one gets different relative contributions from these two components.  In the
model with uniformly heated polar caps, the nonthermal component 
described as a PL
of photon index $\gamma=2.7-2.9$ provides about 80\% in the X-ray flux and
dominates at energies below 0.6 keV and above 2.7~keV.  However, this model
encounters the same problems in the EUV and optical bands as the broken PL 
(the steeper slope of this PL component predicts even higher fluxes in the
$B$ and $V$ bands).

If we assume a more plausible polar cap model with 
temperature decreasing outwards
from the cap center, then the thermal component becomes dominant between 0.06
keV and 2.5 keV, providing some 75\% of the X-ray flux, while the PL component
of $\gamma=1.6-2.5$ dominates outside this band.  In addition to a more 
realistic 
temperature distribution, the latter model is well consistent with the
\euv\/ data and yields estimates on the hydrogen column density in agreement with
the indirect measurements.  As this PL component is fainter than in the two
other models, its extension to the optical falls below the observed radiation
of the white dwarf companion for photon indices $\gamma < 1.9$.  This allows us
to predict that the PL component should be 
observable in the UV
(particularly, far-UV) range where it is brighter than the Wien tail of the
white dwarf spectrum, assuming there is no turnover of the nonthermal spectrum
between the UV and soft-X-ray energies.  On the other side of the X-ray band,
extrapolation of the PL component to the gamma-ray energies above 100~MeV 
predicts a photon flux $f<2\times 10^{-8}$~s$^{-1}$~cm$^{-2}$ ($\gamma>1.6$),
below the upper limit, $f < 1.5\times 10^{-7}$~s$^{-1}$~cm$^{-2}$,
obtained from the {\sl CGRO} EGRET observations (Fierro et al.~1995).

We emphasize that these  models require the thermal radiation to be emitted
from hydrogen (or helium) NS atmosphere. The high spectral resolution of the
ACIS data rules out an atmosphere comprised of heavier chemical elements.

The HRC-S observation of \psr\ has demonstrated the \chan\/ timing capability
at a millisecond level.  The HRC-S pulse profile looks narrower, and the
pulsed fraction is somewhat higher, than those obtained in the earlier \ros\/ 
and \euv\/ observations at lower energies, which could be explained by the
properties of the thermal radiation from polar caps covered with a hydrogen or
helium atmosphere.  On the other hand, the shape of the profile is clearly
asymmetric, with a longer rise and faster decay, which cannot be explained
by a simple axisymmetric temperature distribution. Relativistic effects
(particularly, the Doppler boost) should lead to a different asymmetry ---
a faster rise and longer trail 
(Braje \& Romani 2000; Ford 2000).
The analysis of
HRC-S data demonstrates, for the first time, that the phase of the X-ray
pulse virtually coincides with that of the radio pulse.  If, as we
suggest, the main contribution to the HRC-S band is due to the thermal polar cap
radiation, and if the pulsar radio beam is directed 
along the magnetic axis,
then the radio emission 
must be generated close to
the NS surface --- e.g., the time difference of  $<0.1$ ms between the
X-ray and radio phases corresponds to a distance of $<30$ km, much smaller
than the light cylinder radius, $R_{\rm lc}=275$ km.
Alternatively, if the radio emission is generated at a higher
altitude, the combination of field-line sweepback and aberration
must contrive to cancel the radial travel-time difference.

The \chan\/ observations show no sign of an X-ray PWN that
could accompany the bow-shock revealed by the H$_\alpha$ observations.
Three-sigma upper limits on the PWN brightness (in counts~arcsec$^{-2}$)
can be estimated as $3(b/A)^{1/2}$, where $b$ is the background
surface brightness ($b=0.51$ and 0.28 counts~arcsec$^{-2}$ for the ACIS-S
and HRC-I images, respectively), and $A$ is the 
PWN area 
(we will scale it as $A=1000 f_A$ arcsec$^2$, assuming that a typical transverse
size of the PWN somewhat exceeds the stand-off distance, 
$10''$, of the bow shock). 
For a power-law PWN spectrum with a photon index $\gamma=1.5$--2
(similar to those observed from other PWNe), these upper
limits correspond to the PWN intensities 
$I_{x,{\rm pwn}}<(1.3$--$1.8)\times 10^{-17} f_A^{-1/2}$
and $I_{x,{\rm pwn}}<(3.6$--$5.7)\times 10^{-17} f_A^{-1/2}$ 
erg cm$^{-2}$ s$^{-1}$ arcsec$^{-2}$,
for the ACIS-S and HRC-I, respectively, 
in the 0.1--10 keV range. 
The corresponding upper limits on the PWN X-ray luminosity,
$L_{x,{\rm pwn}}\approx 4\pi d^2 A I_{x,{\rm pwn}}$ are much smaller than the
rotational energy loss rate, $\dot{E}=3.8\times 10^{33}$ erg s$^{-1}$ --- e.g.,
$L_{x,{\rm pwn}} < (3.0$--$4.2)\times 10^{28} f_A^{1/2}$ erg s$^{-1}$ 
for the more sensitive ACIS-S limit.

The low upper limits on the PWN luminosity in X-rays can be simply explained
by a low magnetic field 
in the PWN region, expected for a particle-dominated pulsar wind.
The shock in the
relativistic pulsar wind should be located just interior to
the observed H$_\alpha$ bow shock (Arons \& Tavani 1993).
When the wind electrons
pass through the shock, their
directions of motion
become ``randomized'', and their synchrotron radiation may
result in an X-ray nebula, provided the electron energies and the
magnetic field are high enough in the post-shock region.
The pre-shock magnetic field can be estimated as 
$B_1=[\dot{E}/(f_\Omega r_s^2c)]^{1/2}
[\sigma/(1+\sigma)]^{1/2}=18\,f_\Omega^{-1/2} [\sigma/(1+\sigma)]^{1/2}~\mu$G,
where $r_s\approx 2\times 10^{16}$ cm is the stand-off distance
corresponding to $10''$ at $d=140$ pc, $f_\Omega=\Delta\Omega/(4\pi)\leq 1$
is the collimation factor of the wind,
and  the ``magnetization parameter''
$\sigma$ is the ratio of the Poynting flux to the kinetic energy flux.
The maximum value
of the post-shock magnetic field, 
$B\simeq B_1\simeq 18\, f_\Omega^{-1/2}~\mu$G, is
obtained for $\sigma\gg 1$. However,
according to Kennel \& Coronity (1984; KC84 hereafter), 
a significant fraction of the total
energy flux upstream can be converted into (observable) synchrotron luminosity
downstream only if $\sigma \lapr 0.1$ (e.g., these authors estimate
$\sigma\approx 0.003$ for the Crab pulsar). 
For $\sigma\ll 1$, the post-shock magnetic field
is $B\simeq 3(1-4\sigma)
B_1\simeq 53\,f_\Omega^{-1/2}
\sigma^{1/2}(1-4.5\sigma)~\mu$G
(e.g., from eqs.\ [4.8] and [4.11] of KC84,
$B$ increases from 3\,$\mu$G to 12\,$\mu$G
when $\sigma$ increases from 0.003 to 0.1, at $f_\Omega=1$).
Such low magnetic fields in the shocked wind strongly limit the maximum
energy, $m_ec^2 \Gamma_{\rm max}$, of radiating electrons and,
consequently, the maximum frequency $\nu_{\rm max}$ of the synchrotron
radiation.
Since the Larmor radius of most energetic electrons,
$r_{L}=1.7\times 10^8\, \Gamma_{\rm max} B_{-5}^{~~-1}~{\rm cm}$,
cannot exceed the shock radius $r_s$ substantially, 
we obtain $\Gamma_{\rm max} < 10^8\, f_s B_{-5}$,
$h\nu_{\rm max} \sim (heB/4\pi m_e c)\Gamma_{\rm max}^2 
< 0.6\, f_s^2 B_{-5}^{~~3}$ keV, where $B_{-5}=B/(10\, \mu{\rm G}$),
$f_s\equiv r_L/r_s\sim 1$.
If, for instance, $B<5 f_s^{-2/3}~\mu$G 
(i.e., $\sigma < 0.01\, f_s^{-4/3}$ in the
KC84 model),
the synchrotron emission 
at the bow shock is not expected to be seen in X-rays.

Despite the superior quality of the \chan\/ data, which have allowed us to
detect the hard tail of the pulsar's spectrum and 
pinpoint the absolute phase of the
X-ray pulse, there still remain some open problems.  Although our analysis
strongly favors the thermal+nonthermal interpretation, it is still unclear 
which of the two components dominates in the X-ray radiation of \psr.  To
establish the relative contributions of these components, energy-resolved
timing and time-resolved spectral analysis are needed, which, hopefully, will
be possible with the forthcoming {\sl XMM-Newton} data.  

\acknowledgements
We thank Leisa Townsley for the advise on the ACIS data reduction,
Allyn Tennant for the discussion of the HRC timing issues,
and Willem van Straten for providing the timing ephemeris
given in Table 1.
We also thank Vadim Burwitz for helping in preparation of
the color image.
This work was partially supported by NASA through grants NAG5-7017, NAG5-10865,
and SAO GO0-1126X. 

{}

\newpage
\clearpage

\voffset=-0.4truein
\begin{table}
\begin{center}
\caption{\centerline{Ephemeris parameters for the binary pulsar 
J0437--4715\tablenotemark{a}}}
\begin{tabular}{ll}
\tableline\tableline
Parameter                           & ~~~Value \\
\tableline
R.A.~(J2000)..........................................
& $04^{\rm h}37^{\rm m}15\fs 78651377$~~~~~~ \\
Dec.~(J2000)..........................................
& $-47^\circ 15'~8\farcs46158135$~~~~~~~~~~~~ \\
$\mu_{\alpha}$~(mas/yr)............................................
& $121.4565$ \\
$\mu_{\delta}$~(mas/yr)............................................
& $-71.4494$ \\
Parallax (mas)........................................
& 6.9830 \\
$F$~(Hz)....................................................
& 173.68794899909872242  \\
$\dot{F}$~(10$^{-15}$~s$^{-2}$)..........................................
& $-1.728325769909$~~~~~~~~~~~~~~~         \\
$\ddot{F}$~(10$^{-30}$~s$^{-3}$)..........................................
& $132.1214177$ \\
Epoch of period and position (MJD).....
& 51,194.0~~~~~~~~~~~~~~~~~~~~   \\
DM (cm$^{-3}$~pc)........................................
& 2.64690 \\
Binary period, $P_{\rm b}$~(days)........................ 
& 5.741044490290~~~~~~~~~ \\
$x=a_{\rm p}\sin{i}$~(lt-s)\tablenotemark{b}......................................
& 3.366693427~~~~~~~~~~~~~~~ \\
Eccentricity, $e$........................................
& 0.000019193~~~~~~~~~~~~~~ \\
Longitude of periastron, $\omega$~(deg)........... 
& 1.848063~~~~~~~~~~~~~~~~~~~       \\
Epoch of periastron (MJD)....................    
& 51,194.634254236~~~~~ \\
$\dot{x}$~($10^{-12}$~lt-s~s$^{-1}$).....................................     
& 0.078270~~~~~~~~~~~~~~~~~~~~~~~  \\
$\dot{P_{\rm b}}$~($10^{-12}$~s~s$^{-1}$)....................................... 
& 3.397~~~~~~~~~~~~~~~~~~~~~~~~~~  \\
\tableline
\end{tabular}
\tablenotetext{a}{Based on radio observations made using the Parkes
64-m telescope of the Australia Telescope National Facility
(van Straten et al.~2001)}
\tablenotetext{b}{Projected semi-major axis of the pulsar's orbit}
\end{center}
\end{table}

\clearpage
\newpage
\begin{figure}
 \centerline{\psfig{figure=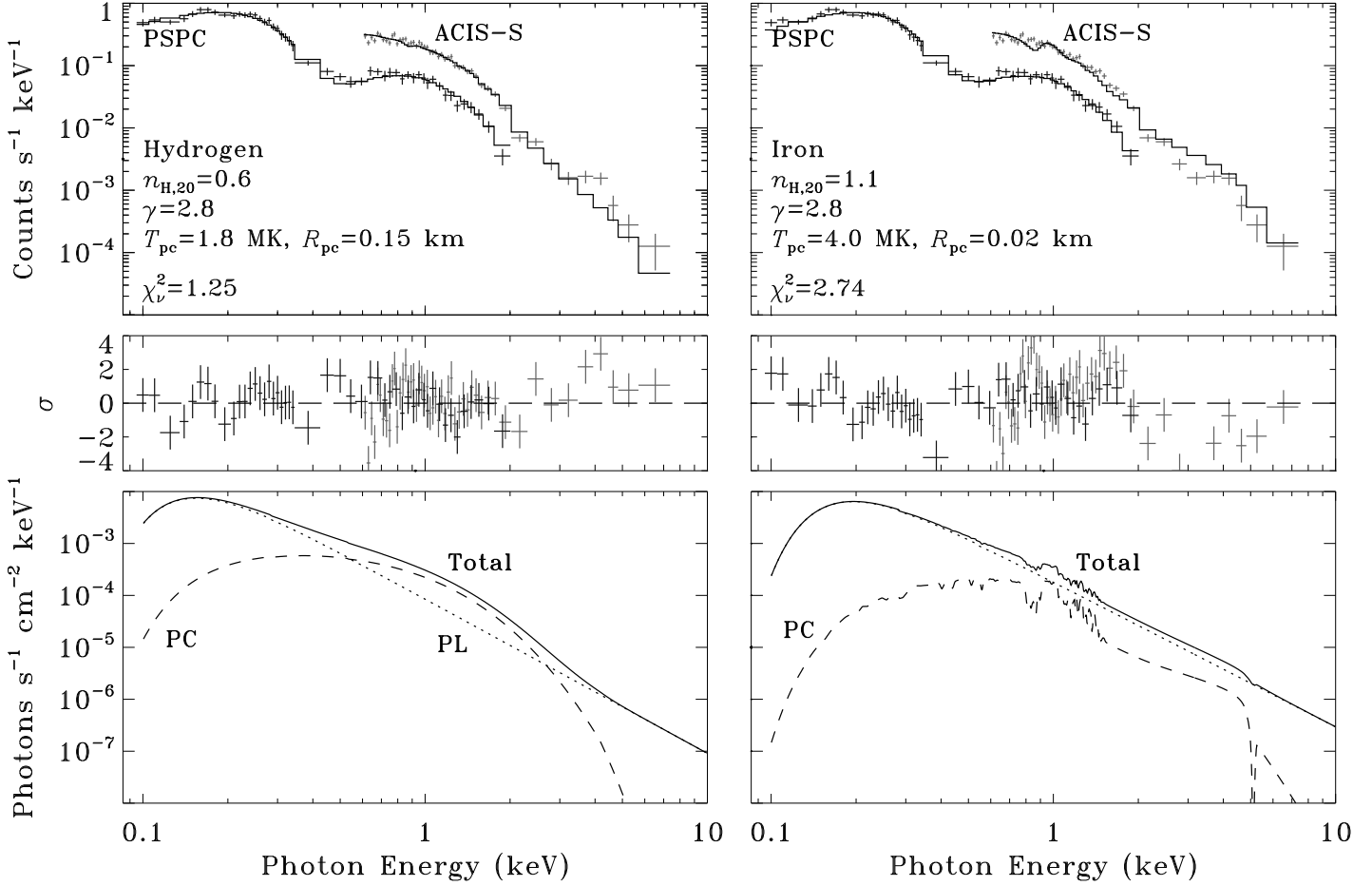,height=12cm}}
 \caption[]{
Power-law plus one-temperature polar cap model fits to the
\ros\/ and \chan\/ spectra.
The left and right panels correspond to polar caps covered
with hydrogen and iron atmospheres, respectively.
}
\end{figure}

\clearpage

\begin{figure}
 \centerline{\psfig{figure=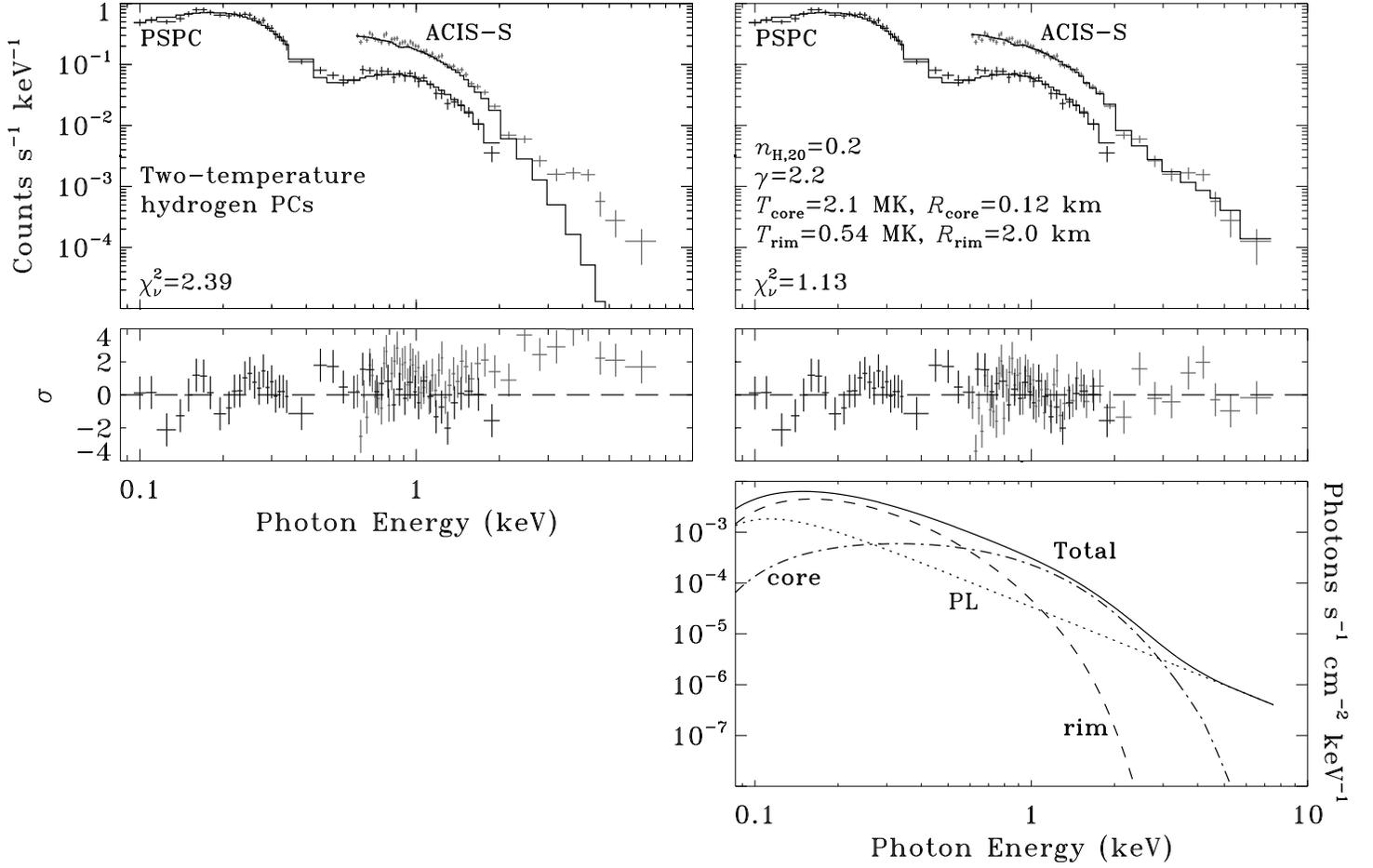,height=12cm}}
 \caption[]{
Two-temperature polar cap model fits with 
and without 
the power-law component (right and left panels, respectively).
}
\end{figure}
\clearpage

\begin{figure}
 \centerline{\psfig{figure=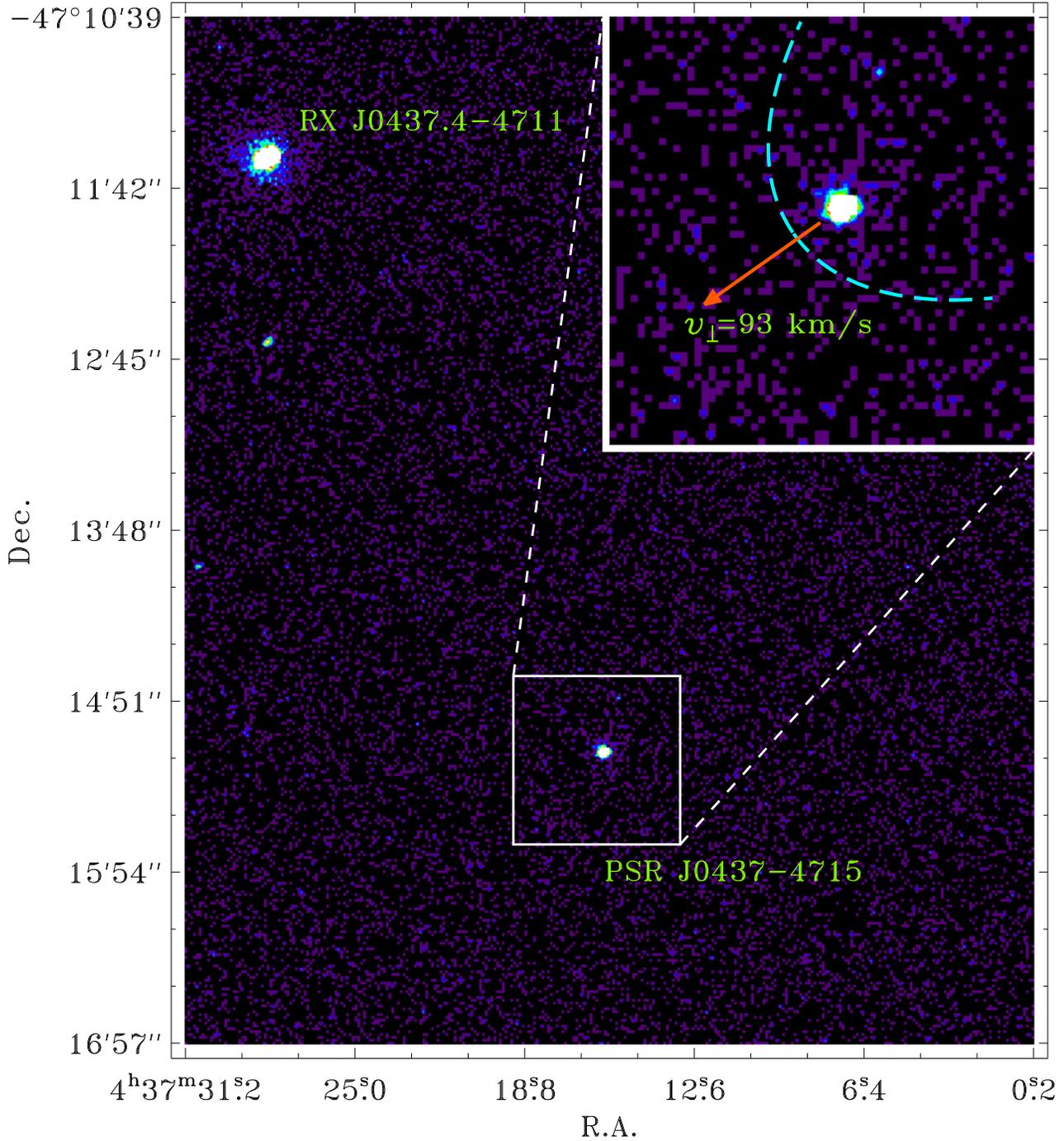,height=20cm}}
 \caption[]{
$5\farcm2\times 6\farcm3$ region of the HRC-I image.
The bright source in the upper-left corner is an AGN
(Halpern \& Marshall 1996).
The inset in the upper-right corner shows a zoomed
$63''\times 63''$
field around the pulsar.
The dashed curve marks the leading edge of
the H$_\alpha$ bow-shock. The arrow shows the direction of the
pulsar's proper
motion, with the transverse velocity $v_\perp$.
}
\end{figure}
\clearpage

\begin{figure}
 \centerline{\psfig{figure=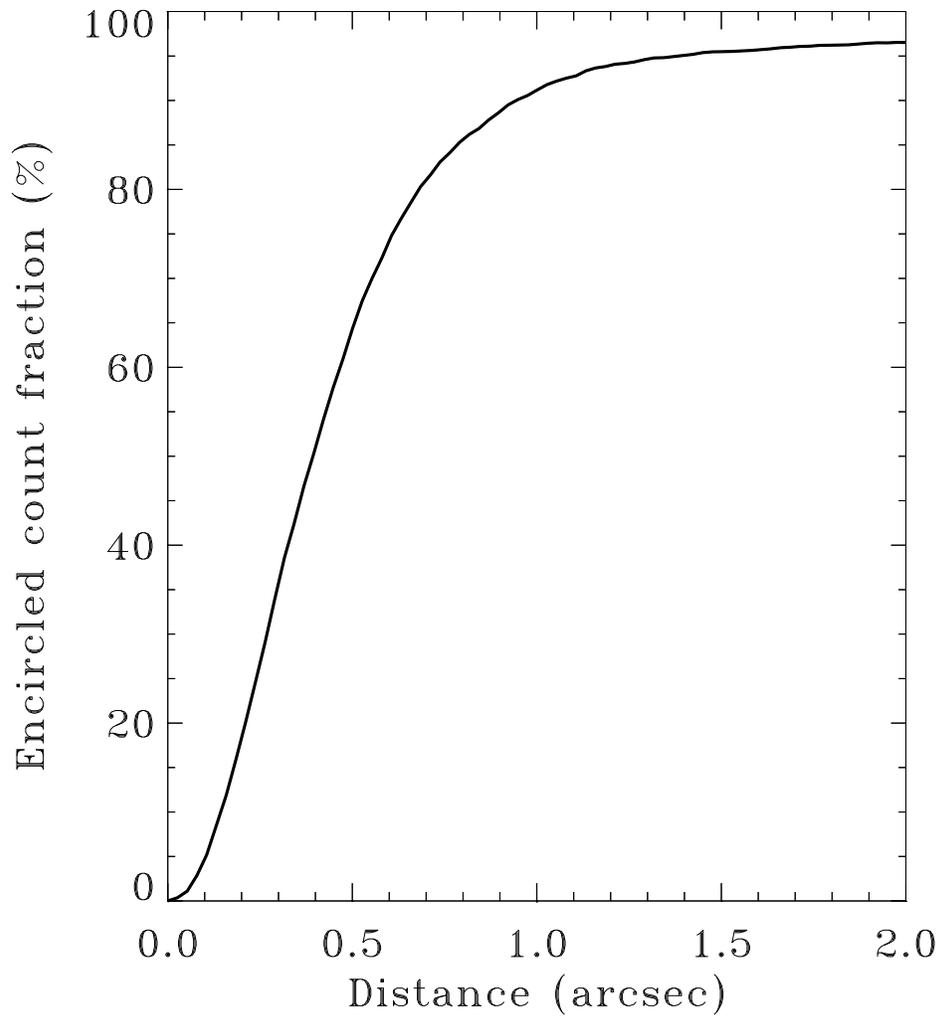,height=12cm}}
 \caption[]{
Radial distribution of encircled fraction of pulsar counts in the HRC-I data.
}
\end{figure}
\clearpage

\begin{figure}
 \centerline{\psfig{figure=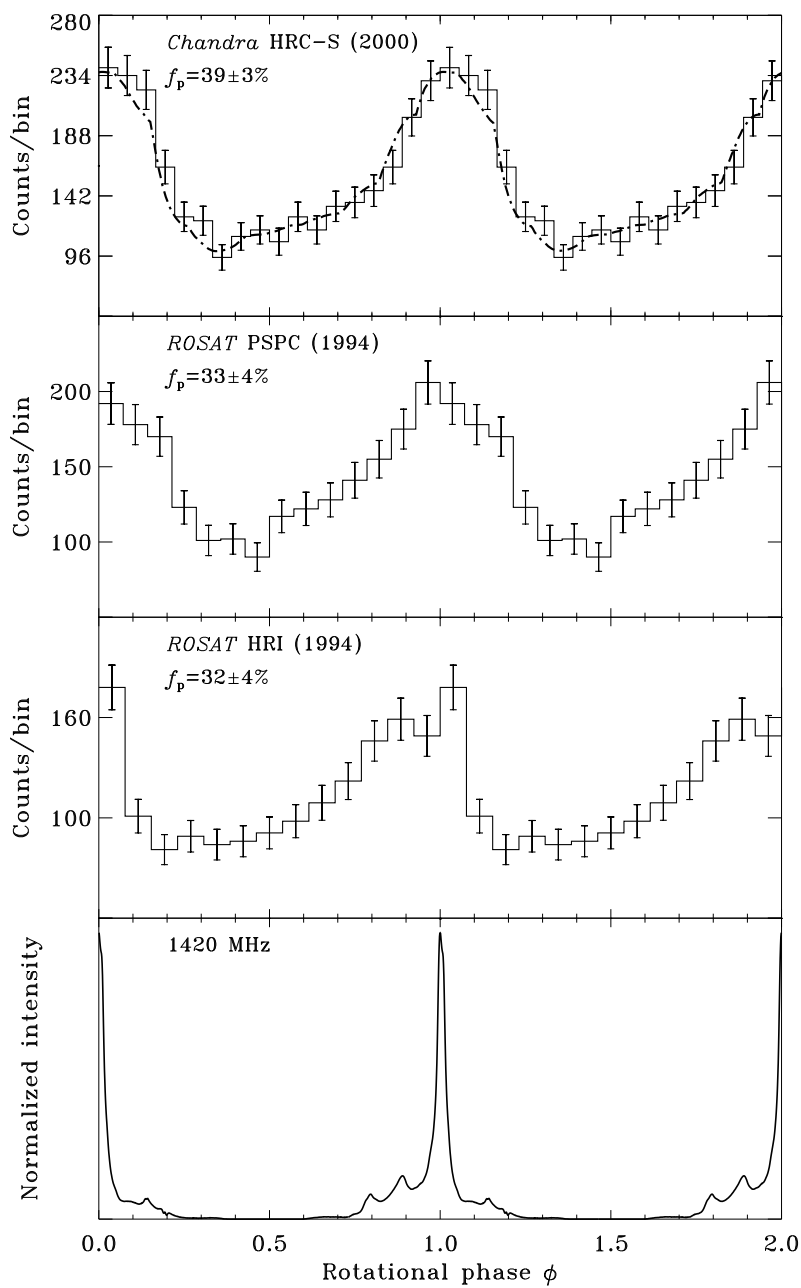,height=20cm}}
 \caption[]{
Pulse profiles of \psr\ from the \chan, \ros,
and radio observations.
The smooth light curve in the upper panel is obtained by
averaging of the 20-bin profile over the starting phase.
The zero phase corresponds to the peak of the radio pulse
shown in the lower panel which is the summed profile from
the 1400 MHz radio observations made with the Caltech
correlator at Parkes. The absolute phases of the {\sl ROSAT} pulses
were not measured; we chose the phases to approximately co-align these pulses
with the {\sl Chandra} pulse.
}
\end{figure}

\end{document}